\documentclass[review]{elsarticle}%
\usepackage{lineno,hyperref}
\usepackage{amsmath}
\usepackage{amsfonts}
\usepackage{amssymb}
\usepackage{graphicx}%
\providecommand{\U}[1]{\protect\rule{.1in}{.1in}}
\modulolinenumbers[5]
\journal{Journal of \LaTeX\ Templates}
\begin{document}
\begin{frontmatter}
\title{Magnonics vs. Ferronics}
\author[AIMR,IMR,CSRN,Groningen]{Gerrit E.W. Bauer}
\author[AIMR]{Ping Tang}
\author[NIMS]{Ryo Iguchi}
\author[NIMS,IMR,CSRN]{Ken-ichi Uchida}
\address[AIMR]{WPI-AIMR, Tohoku
University, 2-1-1 Katahira, 980-8577 Sendai, Japan}
\address[IMR]{IMR, Tohoku
University, 2-1-1 Katahira, 980-8577 Sendai, Japan}
\address[CSRN]{Center for Spintronics Research Network, Tohoku University, Sendai 980-8577, Japan}
\address[Groningen]{Zernike Institute for Advanced Materials, University of Groningen, 9747 AG Groningen, Netherlands}
\address[NIMS]{National Institute for Materials Science, Tsukuba 305-0047, Japan}
\begin{abstract}
Magnons are the elementary excitations of the magnetic order that carry spin, momentum, and energy.
Here we compare the magnon with the \textit{ferron}, i.e. the elementary excitation of the electric dipolar order,
that transports polarization and heat in ferroelectrics.
\end{abstract}
\end{frontmatter}

\linenumbers

\section{Introduction}

Much of condensed matter physics addresses weak excitations of materials and
devices from their ground states. More often than not the complications caused
by electron correlations can be captured by the concept of approximately
non-interacting quasi-particles with well-defined dispersion relations.
Quasi-particles are not eigenstates and eventually decay on characteristic
time scales that depend on material and environment. In conventional electric
insulators, the lowest energy excitations are lattice waves with associated
bosonic quasi-particles called phonons. The elementary excitations of magnetic
order are spin waves and their bosonic quasi-particles are the
\textquotedblleft magnons\textquotedblright\ \cite{Rezende20}. In magnetic
insulators, magnons and phonons coexist in the same phase space and can form
hybrid quasi-particles that may be called \textquotedblleft magnon
polarons\textquotedblright. Yttrium iron garnet (YIG) is the material of
choice to study magnons, phonons, and magnon polarons because of their
longevity in YIG bulk and thin-film single crystals. For more than half a
century, Professor Sergio Rezende and his team seminally contributed to our
understanding of magnons, phonons, and their hybrids by experimental and
theoretical research, see e.g. \cite{Rezende67,Rezende21}. Recently we
extended Prof. Rezende's Boltzmann theory of the spin Seebeck effect
\cite{Rezende14} in magnetic insulators by introducing \textquotedblleft
ferrons\textquotedblright, i.e. the excitation of the electric dipolar order
in ferroelectrics (FEs), and explored their ability to transport heat and
polarization, including the associated Seebeck and Peltier effects
\cite{Bauer2021,Tang2021}. Here we expound the analogies and differences
between magnets and FEs in their ground states \cite{Spaldin}\ as well as in
their dynamic and transport properties, thereby comparing the magnonics of
ferromagnetic insulators with the \textquotedblleft
ferronics\textquotedblright\ of ferroelectric insulators.

\section{Magnetism vs. ferroelectricity}

\subsection{Dipoles}

A magnetic dipole $\mathbf{M}$ is called \textquotedblleft
Amp\`{e}rian\textquotedblright\ \cite{Ampere} because it generates magnetic
fields by the intrinsic or orbital angular momentum of charged particles,
implying broken time reversal symmetry. The interference with an external
magnetic field $\mathbf{H}$ causes the Zeeman interaction
\begin{equation}
E_{M}=-\mathbf{M}\cdot\mu_{0}\mathbf{H}%
\end{equation}
with vacuum permeability $\mu_{0}$ that leads to the Landau-Lifshitz equation
of motion.
\begin{equation}
\mathbf{\dot{M}}=-\gamma\mathbf{M}\times\mathbf{H,}%
\end{equation}
where $\gamma=g_{e}\mu_{B}/\hbar$ is the modulus of the gyromagnetic ratio,
$g_{e}$ is the electron g-factor and $\hbar=h/\left(  2\pi\right)  $ Planck's
reduced constant.

Two local spins $\mathbf{S}_{1}$ and $\mathbf{S}_{2}$ with magnetic moments
$\mathbf{M}_{i}=-\gamma\mathbf{S}_{i}$ can couple by the exchange interaction
\begin{equation}
E_{x}=-J\mathbf{S}_{1}\cdot\mathbf{S}_{2},
\end{equation}
where the exchange integral $J$ vanishes exponentially as a function of
distance between the local moments on an interatomic length scale. The
magneto- and electro-dipolar interactions have the same angle and distance
dependence, but different prefactors \cite{Chandra07}. The interaction energy
of two parallel magnetic moments with $\mathbf{M}_{i}=\mu_{B}$ at a distance
$r$ is
\begin{equation}
F_{M}=\frac{\mu_{0}\mu_{B}^{2}}{4\pi}\frac{1}{r^{3}}\sim\frac{\alpha^{2}}%
{4\pi}\left(  \frac{a_{B}}{r}\right)  ^{3},
\end{equation}
where $\alpha=1/137$ is the fine structure constant, $\mu_{B}$ the Bohr
magneton, and $a_{B}$ the Bohr radius.

An electric dipole $\mathbf{P}$ is called \textquotedblleft
Gilbertian\textquotedblright\ \cite{Gilbert}, i.e. a directional difference of
positive and negative charges. $\mathbf{P}$ breaks inversion symmetry, but it
is time-reversal invariant. A dipole interacts with an electric field
$\mathbf{E}$
\begin{equation}
E_{P}=-\mathbf{P}\cdot\mathbf{E.}%
\end{equation}
A point electric dipole is not affected by a constant electric field, so
$\mathbf{\dot{P}}=\mathbf{0.}$

There is no such thing as an exchange interaction between electric dipoles, so
they interact only electrostatically. The coupling of two parallel electric
dipoles with $P\sim ea_{B}\sim10^{-29}%
\operatorname{C}%
\operatorname{m}%
$ scales like
\begin{equation}
F_{P}=\frac{P^{2}}{4\pi\epsilon_{r}\epsilon_{0}}\frac{1}{r^{3}}\sim\frac
{1}{4\pi\epsilon_{r}}\left(  \frac{a_{B}}{r}\right)  ^{3},
\end{equation}
where $\epsilon_{r}$ is the relative dielectric constant of the medium that
embeds the dipoles. Therefore
\begin{equation}
\frac{F_{P}}{F_{M}}=\frac{1}{\alpha^{2}\epsilon_{r}}\gg1. \label{PMratio}%
\end{equation}

\subsection{Ground states}

\textit{Magnets - }The magnetic order in electric insulators can be described
by the Heisenberg model of local spins coupled by the exchange interaction $J$
between nearest neighbors (\textit{n.n.})
\begin{equation}
H=-\frac{J}{2}\sum_{ij}^{n.n.}\mathbf{S}_{i}\cdot\mathbf{S}_{j}.
\end{equation}
For a positive $J$ the ground state is ferromagnetic with a coarse-grained
continuous field $\mathbf{m}M_{0},$ where $M_{0}=\left\vert \gamma
\mathbf{S}/\Omega\right\vert $ is the magnetization and $\Omega$ is the volume
occupied by a single spin. The direction and position dependent unit
vector\ $\mathbf{m}$ of the ground state minimizes the magnetic energy that
includes the exchange energy and the Zeeman interaction with applied and
effective magnetic fields.

\textit{FEs - }Ferroelectricity is the ordered state of a large number of
microscopic electric dipoles with an associated permanent electric
polarization. Roughly two types of FEs can be distinguished, viz.
\textquotedblleft displacive\textquotedblright\ and \textquotedblleft
order-disorder\textquotedblright. In the former, the order emerges during a
soft-phonon structural phase transition that breaks inversion symmetry of the
unit cell, while in the latter stable molecular dipoles order during the phase
transition. We can write the unit cell dipole as $\mathbf{P}_{0}%
=Q\boldsymbol{\delta}$, where $Q$ is the ionic charge and $\boldsymbol{\delta
}$ a displacement vector. The polarization density $p_{0}=P_{0}/\Omega$, where
$\Omega$ is the unit cell volume.

FEs have a large dielectric constant $\epsilon_{r}\sim10^{3},$ so according to
Eq. (\ref{PMratio}) $F_{P}/F_{M}\sim20$. Since the magnetostatic interaction
governed by $F_{M}$ are typically in the GHz regime, the electric dipolar
interactions are still much smaller than ambient thermal energies. Since there
is no exchange interaction that order the dipoles, ferroelectricity appears to
be a consequence, but not the origin of the phase transition to the
symmetry-broken ground state. However, the dipolar interactions affect the
polarization direction and texture including domain formation in larger
structures stronger than in ferromagnets.

\subsection{Excitations}

Magnonics and ferronics address the small-amplitude dynamics of the dipolar
order at temperatures below the phase transition.

\textit{Magnets -} In magnets, the low energy excitations are spin waves, i.e.
plane-wave-like modulated precessions around the equilibrium magnetization
with a time and position dependent phase. The exchange energy contribution
$\omega_{k}^{\left(  \mathrm{x}\right)  }\sim Jk^{2}$ vanishes at small wave
numbers $k.$ In this limit the magnetodipolar interaction with $\omega
_{k}^{\left(  \mathrm{dip}\right)  }-\omega_{0}^{\left(  \mathrm{dip}\right)
}\sim k$ dominates. Spin and lattice waves coexist in the same regions of
reciprocal space. When their coupling by the magnetic anisotropy and
magnetoelasticity is larger than the level splitting and broadening, they form
hybrid states or magnon polarons that transport magnetization with the sound
velocity \cite{Flebus}. The interaction between spins and lattice is
relativistic and in general weak, so in most magnets magnon polarons form, if
at all, only in relatively small volumes of phase space.

\textit{FEs - }The electric dipoles in FEs are defined by the coordinates of
the charged ions that have mass and are subject to electric fields and elastic
forces. In contrast to magnets, the polarization dynamics for $k\rightarrow0$
is governed by transverse optical phonons in the THz regime. The dipolar
contribution to the restoring force constant can be estimated for a film
geometry as
\begin{equation}
C_{P}=\frac{Q^{2}}{\epsilon_{r}\epsilon_{0}\Omega}=\mathcal{O}\left(
10^{2}\,\mathrm{J/m^{2}}\right)  ,
\end{equation}
which is much smaller than typical elastic ones. To a good approximation the
elementary excitations of the FE order therefore appear to be mechanical, i.e.
electrically polarized phonons rather than oscillating massless dipoles, which
are solutions of a lattice dynamics problem.

\section{Transport}

Transport of magnetic order in the form of spin currents is crucial in
spintronics and spin caloritronics \cite{Bauer}. Surprisingly, the transport
of the FE order has attracted little attention, see bibliography in
\cite{Bauer2021}. Here we review our formulation of polarization transport in
FEs that is inspired by the models for spin transport in magnetic insulators
\cite{Rezende14,Cornelissen,Flebus,Meier}.

\subsection{Linear response}

\textit{Magnets - }Magnons in magnetic insulators carry energy and spin
currents when subject to temperature and magnetic field gradients $\partial T$
and $\partial H$. In the linear response regime of a homogeneous and isotropic
magnet, the extended \textquotedblleft Ohm's Law\textquotedblright\ reads%
\begin{equation}
\left(
\begin{array}
[c]{c}%
-j_{m}\\
j_{q}%
\end{array}
\right)  =\sigma_{m}\left(
\begin{array}
[c]{cc}%
1 & S_{m}\\
\Pi_{m} & \kappa/\sigma_{m}%
\end{array}
\right)  \left(
\begin{array}
[c]{c}%
\partial H\\
-\partial T
\end{array}
\right)  , \label{Onsagerm}%
\end{equation}
where $j_{m}$ $(-j_{m})$ is the magnetization (magnon) current density defined
by\ the conservation law $\partial j_{m}=-\dot{M}$, $j_{q}$ is the heat
current density, $\sigma_{m}$ is a magnetization or magnon conductivity,
$\kappa$ the thermal conductivity, while $\Pi_{m}$ and $S_{m}=\Pi_{m}T$ \ are
the spin Peltier and Seebeck coefficients, respectively. The signs are chosen
such that the transport parameters are positive for a simple ferromagnet.

\textit{FEs - }Similar equations govern electric polarization transport, but
$H$ is replaced by an electric field $E$:%
\begin{equation}
\left(
\begin{array}
[c]{c}%
-j_{p}\\
j_{q}%
\end{array}
\right)  =\sigma_{p}\left(
\begin{array}
[c]{cc}%
1 & S_{p}\\
\Pi_{p} & \kappa/\sigma_{p}%
\end{array}
\right)  \left(
\begin{array}
[c]{c}%
\partial E\\
-\partial T
\end{array}
\right)  , \label{Onsagerp}%
\end{equation}
and $j_{p}$ ($-j_{p})$ is the polarization (ferron) current density defined
by\ the conservation law $\partial j_{p}=-\dot{P}$, $\sigma_{p}$ is a
polarization or ferron conductivity, $\Pi_{p}$ and $S_{p}=\Pi_{p}T$ are the
polarization Peltier and Seebeck coefficients, respectively. The signs are
chosen such that the transport parameters are positive for simple FEs. The
polarization current should be distinguished from the displacement current
$\dot{P},$ which is not a transport property.

\subsection{Diffusion theory}

\textit{Magnets - }In magnetic insulators, the diffusion picture of transport
has successfully made contact with experiments with only a small number of
adjustable parameters. At equilibrium with field $H_{0}$ and temperature
$T_{0}$ the magnons are distributed according to the Bose-Einstein function
with zero chemical potential (or Planck's function)
\begin{equation}
f_{BE}^{\left(  0\right)  }\left(  \varepsilon\right)  =\left[  \exp\left(
\frac{\varepsilon+\hbar\gamma H_{0}}{k_{B}T_{0}}\right)  -1\right]  ^{-1}.
\end{equation}
The crucial assumption is approximate local equilibration of the magnon
spectral distribution to the form \cite{Duine}
\begin{equation}
f_{BE}\left(  x,\varepsilon\right)  =\left[  \exp\left(  \frac{\varepsilon
+\hbar\gamma H\left(  x\right)  -\mu_{m}\left(  x\right)  }{k_{B}T_{m}\left(
x\right)  }\right)  -1\right]  ^{-1} \label{fBE}%
\end{equation}
in terms of a slowly varying magnetic field $H\left(  x\right)  =H_{0}%
+\triangle H\left(  x\right)  ,$ non-equilibrium magnon chemical potential or
magnon accumulation $\mu_{m}\left(  x\right)  $ and temperature $T_{m}\left(
x\right)  =T_{0}+\triangle T_{m}\left(  x\right)  $. In magnetic insulators,
this assumption can be justified by efficient magnon-conserving magnon-phonon
and magnon-magnon scattering, while magnon-non-conserving damping processes
are weak \cite{Cornelissen}. The magnon number density is then reduced to%
\begin{equation}
-m\left(  x\right)  =\int f_{BE}\left(  x,\varepsilon\right)  \rho_{m}\left(
\varepsilon\right)  \frac{d\varepsilon}{2\pi}, \label{mx}%
\end{equation}
where $\rho_{m}\left(  \varepsilon\right)  $ is the magnon energy density of
states. In linear response and otherwise homogeneous systems%
\begin{align}
f_{BE}\left(  x,\varepsilon\right)   &  =f_{BE}^{\left(  0\right)  }\left[
1-f_{BE}^{\left(  0\right)  }\exp\left(  -\frac{\varepsilon+\hbar\gamma H_{0}%
}{k_{B}T_{0}}\right)  \right. \nonumber\\
&  \left.  \left(  \frac{\hbar\gamma\triangle H-\mu_{m}}{k_{B}T_{0}}%
+\frac{\left(  \varepsilon+\hbar\gamma H_{0}\right)  \triangle T}{k_{B}%
T_{0}^{2}}\right)  \right]  \label{fBEexp}%
\end{align}
and
\begin{equation}
m\left(  x\right)  -m_{0}=\chi_{m}\left(  \triangle H-\frac{\mu_{m}}%
{\hbar\gamma}\right)  +\chi_{T}\triangle T_{m},
\end{equation}
where $m_{0}$ is the thermal equilibrium value. The susceptibilities%
\begin{align}
k_{B}T_{0}\chi_{m}  &  =-\int\left(  f_{BE}^{\left(  0\right)  }\right)
^{2}e^{-\frac{\varepsilon+\hbar\gamma H_{0}}{k_{B}T_{0}}}\rho_{m}%
\frac{d\varepsilon}{2\pi}\\
k_{B}T_{0}^{2}\chi_{T}  &  =-\int\left(  f_{BE}^{\left(  0\right)  }\right)
^{2}e^{-\frac{\varepsilon+\hbar\gamma H_{0}}{k_{B}T_{0}}}\left(
\varepsilon+\hbar\gamma H_{0}\right)  \rho_{m}\frac{d\varepsilon}{2\pi}%
\end{align}
parameterize the response to \emph{constant} field and temperature changes and
are crucial parameters for magnetocaloric effects. The finite lifetime
$\tau_{m}\left(  \tau_{q}\right)  $ of the magnon (energy) density modifies
the conservation relations%
\begin{align}
-\partial j_{m}  &  =\dot{m}+\frac{m-m_{0}}{\tau_{m}}\label{consvm}\\
-\partial j_{q}  &  =\dot{q}+\frac{q-q_{0}}{\tau_{q}}.
\end{align}
where the subscript $0$ indicates the equilibrated values. Efficient
magnon-conserving magnon-phonon interactions render a short $\tau_{q}$ and the
lattice has a large heat capacity. Then $T=T_{0}$, i.e. the magnon temperature
equals the lattice temperature everywhere. This approximation should be
reconsidered in the low temperature regime \cite{Elyasi}. Combining Eqs.
(\ref{Onsagerm},\ref{mx}-\ref{consvm}) and assuming $\partial^{2}H=0$ we find
for the steady state%
\begin{equation}
\partial^{2}\mu_{m}=\frac{\mu_{m}}{\ell_{m}^{2}}.
\end{equation}
with magnon diffusion length $\ell_{m}=\sqrt{\sigma_{m}\tau_{m}/\chi_{m}}.$
The transport parameter $\sigma_{m}$ indicates the realm of spin caloritronics
\cite{Bauer}. We note that the divergence of a spin accumulation is equivalent
to magnetic charges and stray magnetic fields that most studies neglect.

\textit{FEs - }By the same arguments, we arrive at the polarization diffusion
equation for the chemical potential or polarization accumulation $\mu_{p}$
\begin{equation}
\partial^{2}\mu_{p}=\frac{\mu_{p}}{\ell_{p}^{2}}. \label{Diffp}%
\end{equation}
with $\ell_{p}=\sqrt{\sigma_{p}\tau_{p}/\chi_{p}}$. Equation (\ref{Diffp})
relies on assumptions that are well-tested for magnetic insulators at room
temperature, but not for ferroelectrics. The existence of an accumulation
$\mu_{p}$ as a driving force for transport implies that the relaxation time
$\tau_{p}$ is much larger than the scattering life time $\tau_{r}$ that limits
the conductivity $\sigma_{p}.$ In contrast to a magnon accumulation, a
polarization accumulation is not protected in the non-relativistic limit,
however, so the assumption $\tau_{r}\gg\tau_{p}$ should be better justified.
An example of a scattering process that limits transport but conserves
polarization is intermode phonon back scattering by defects.$\tau_{p}$ should
be measurable by the electrocaloric response to small amplitude pulsed or AC
electric fields along the equilibrium polarization. Polarization accumulations
betray their presence with gradients that generate space electric charges and
observable electric fields.

\section{Microscopic theory}

Here we discuss theories that address the material and device-dependent
parameters discussed above. In magnonics, the results compare well with the
parameters fitted to experiments. No experiments are available for
ferroelectrics, however.

\subsection{Excitations}

\textit{Magnets - }The Heisenberg model is the starting point of most
calculations. For large spins such as $S=5/2$ for half-filled 3d-shells,
atomistic simulations give an appropriate picture for the spatiotemporal
dynamics at finite temperatures \cite{Barker}. In the long wave-length limit,
the Landau-Lifshitz Gilbert (LLG) equation
\begin{equation}
\mathbf{\dot{m}=}-\gamma\mathbf{m}\times\mathbf{H}_{\mathrm{eff}}+\alpha
_{G}\mathbf{m}\times\mathbf{\dot{m}} \label{LLG}%
\end{equation}
is well established, in which $\mathbf{H}_{\mathrm{eff}}$ is an effective
magnetic field and $\alpha_{G}$ the Gilbert damping constant. $\mathbf{H}%
_{\mathrm{eff}}$ contains a stochastic term with a correlation function that
in equilibrium obeys the fluctuation-dissipation theorem in terms of the
Gilbert damping and temperature, and can be used to compute the spin-spin
correlation and response functions \cite{Barker}. The linearized
LLG\ equation, equivalent to the Heisenberg equation to lowest order in the
Holstein-Primakoff expansion \cite{Rezende20}, is appropriate for weakly or
thermally excited magnon gases. Its solutions are the magnon dispersion
relations, amplitudes, and group velocities, which are the starting point for
transport theories such as the Boltzmann formalism. The magnetization dynamics
consists of small transverse fluctuations $\mathbf{m}_{\bot}$ that reduce the
net magnetization magnetization $M$ relative to the ground state $M_{0}$
\begin{equation}
-m_{z}=\frac{\mathbf{m}_{\bot}^{2}}{2}. \label{MFE}%
\end{equation}
The (quantum) thermal average $-\overline{m_{z}}$ may be interpreted as the
local number of magnons and $-m=\overline{\mathbf{m}_{\bot}^{2}}M_{0}/\left(
2g\mu_{B}\right)  $ in Eq. (\ref{mx}).

Magnetoelastic interactions mix the magnons with phonons to create hybrid
magnon polarons in reciprocal space that are beyond the scope of
micromagnetics. An appropriate model for small wave numbers is an ensemble of
coupled harmonic oscillators with weak level repulsions at the crossing points
of the magnon and phonon dispersions \cite{Flebus}.

\textit{FEs - }We focus on the symmetry-broken phase at temperatures well
below the phase transition with a finite macroscopic polarization per unit
cell%
\begin{equation}
\mathbf{P}_{0}=\sum_{s}^{\mathrm{unit\,cell}}Q_{s}\mathbf{r}_{s}^{\left(
0\right)  }.
\end{equation}
$\mathbf{r}_{s}^{\left(  0\right)  }$ is the location of the $s$-th ion in a
unit cell with net charge $Q_{s}$ and $\sum_{s}Q_{s}=0.$ We do not address
here the complications caused by surface charges. Finite temperatures or
external excitations induce lattice vibrations that affect the polarization.
The elementary excitation of the crystal is a phonon with wave number
$\mathbf{k}$ in a band $\sigma,$ polarization $\mathbf{e}_{\mathbf{k}\sigma}$
and frequency $\omega_{\mathbf{k}\sigma}$ that modulates the position
$\mathbf{r}_{ls}$ of an ion $s$ with mass $M_{s}$ in the $3N$ unit cells with
index $l=\left\{  1,\cdots,3N\right\}  :$%
\begin{equation}
{\boldsymbol{u}}_{\mathbf{k}\sigma}(l,s)=\frac{1}{\sqrt{NM_{s}}}%
\mathbf{e}_{\mathbf{k}\sigma}(s)e^{i\mathbf{k\cdot r}_{l}-i\omega
_{\mathbf{k}\sigma}t}.
\end{equation}
The local dipole then fluctuates according to%
\begin{equation}
\triangle\mathbf{P}_{l}=\sum_{\mathbf{k}\sigma s}a_{\mathbf{k}\sigma}%
Q_{s}{\boldsymbol{u}}_{\mathbf{k}\sigma}(l,s)
\end{equation}
where $a_{\mathbf{k}\sigma}$ is the phonon amplitude. The polarization
$\mathbf{e}$ has components along and normal to the equilibrium polarization.
The dipolar dynamics consists of a rotation by an angle $\theta$ normal to and
a deformation in the direction $z$ of the dipolar order, which modifies the
projection
\begin{equation}
\left(  \triangle P_{l}\right)  _{z}=\left(  P_{0}+\triangle P_{l}^{\parallel
}\right)  \cos\theta.
\end{equation}
Since harmonic oscillators do not change the average position of the ions and
therefore the polarization, we have to take into account non-linearities. In
magnets, the high energy cost of changing the modulus of the magnetization
leads to the transverse dynamics described by the LLG Eq. (\ref{LLG}). In
\textquotedblleft order-disorder\textquotedblright\ FEs such as KNO$_{3}$ or
NaNO$_{2}$ we have a similar situation, because the nitrate and nitrite
molecular units are characterized by stable permanent dipoles. The
low-frequency phonons are then polarized along the minimum energy path that
switches the polarization without deforming the strongly bound molecular
units. Small transverse fluctuations $\triangle\mathbf{P}_{l}^{\perp}$ reduce
the polarization projection and thereby the macroscopic polarization by
\begin{equation}
\left(  \triangle P_{l}\right)  _{z}=-\frac{\left\vert \triangle\mathbf{P}%
_{l}^{\perp}\right\vert ^{2}}{2P_{0}}+\triangle P_{l}^{\parallel}%
+\mathcal{O}[(\triangle\mathbf{P}_{l}^{\perp})^{2},\triangle P_{l}^{\parallel
}].
\end{equation}
Here the longitudinal fluctuations $\triangle P_{l}^{\parallel}$ are
corrections that are disregarded in the Landau-Lifshitz-Gilbert dynamics of
the magnetization (\ref{MFE}). The transverse fluctuations or \emph{ferrons}
reduce the polarization, just as the magnons reduce the magnetization. This
\textquotedblleft ferron\textquotedblright\ approximation should be accurate
for the order-disorder type as argued above, but it is as yet untested for
displacive FE's.

The averaged polarization is reduced by
\begin{equation}
\overline{\triangle p}=\frac{1}{V}\sum_{\mathbf{k}\sigma}\triangle
p_{\mathbf{k}\sigma}=-\frac{1}{2P_{0}V}\sum_{\mathbf{k}\sigma}\left\vert
a_{\mathbf{k}\sigma}\right\vert ^{2}\left\vert \mathbf{F}_{\mathbf{k}\sigma
}\right\vert ^{2} \label{DelP}%
\end{equation}
where%
\begin{equation}
\mathbf{F}_{\mathbf{k}\sigma}=\sum_{s}\frac{Q_{s}}{\sqrt{M_{s}}}%
\mathbf{e}_{\mathbf{k}\sigma}^{\perp}(s) \label{DelF}%
\end{equation}
After quantizing the oscillators, we arrive at the thermally averaged
polarization per unit cell
\begin{equation}
\overline{\triangle p}=-\frac{\hbar^{2}}{4P_{0}V}\sum_{\mathbf{k}\sigma}%
\frac{\left\vert \mathbf{F}_{\mathbf{k}\sigma}\right\vert ^{2}}{\varepsilon
_{\mathbf{k}\sigma}}f_{BE}^{\left(  0\right)  }\left(  \varepsilon
_{\mathbf{k}\sigma}\right)  \ \label{PEQ}%
\end{equation}
and the susceptibilities that govern the electrocaloric and pyroelectric
susceptibilities $\chi_{p}=\partial\overline{\triangle p}$/$\partial E$ and
$\chi_{T}=\partial\overline{\triangle p}$/$\partial T.$ $-\overline{\triangle
p}/p_{0}$ may be interpreted as an effective number of ferron excitations. In
contrast to the pure magnon case, but similar to that for the magnon polaron,
each quasiparticle excitation contributes with a state-dependent weight to the
reduction of the polarization.

These properties can be computed by conventional lattice dynamics codes for
realistic models. We may capture the essential physics by a one-dimensional
harmonic oscillator model of a diatomic chain of atoms with equal masses $M,$
opposite charges $\pm Q$, and force constants for longitudinal $\left(
C_{L}\right)  $ and transverse $\left(  C_{T}\right)  $ motions. The FE phase
transition shifts the ions in each unit cell\ to generate a permanent electric
dipole $\mathbf{P}_{0}=Q\boldsymbol{\delta}$ with shift vector
$\boldsymbol{\delta}$. The polarization $\mathbf{p}_{0}=\mathbf{P}_{0}/a^{3}$
can point in any direction, but we focus here on a dipolar order along or
normal to the chain. The chain has three phonon branches, one longitudinal and
two transverse modes, in a Brillouin zone with boundaries $\left\vert
k\right\vert \leq\pi/a$. The FE transition doubles the size of the unit cell,
folding the bands at $\left\vert k\right\vert \leq\pi/\left(  2a\right)  $
into acoustic and optical ones. We simplify this already primitive model even
further by assuming that all transverse force constants are the same. The
longitudinal phonons are not ferroelectrically active but contribute to the
heat conductance. We may entirely disregard the high-frequency longitudinal
optical phonon mode. In \cite{Tang2021} we plot the phonon bands for the two
main polarization directions. Remarkably, the polarization of the acoustic
mode can be switched off completely by rotating the direction of the FE order
from in-chain to perpendicular to the chain. This should generate drastic
effects on the caloric and caloritronic properties.

\subsection{Transport}

The transport coefficients can be computed straightforwardly with the above
models for diffuse and ballistic transport. Rezende \textit{et al}.
\cite{Rezende14} and Cornelissen \textit{et al}. \cite{Cornelissen} formulated
a Boltzmann equation in the relaxation time approximation, where the latter
focusses on the role of the magnon chemical potential. Flebus \textit{et
al}.'s \cite{Flebus} spin Seebeck effect theory for magnon polarons can be
easily adapted to handle ferrons. Meier and Loss' \cite{Meier} scattering
theory for magnon transport in ballistic spin chains inspired our ferron
transport formulation \cite{Tang2021}

\subsubsection{Boltzmann equation}

\textit{Magnets - }The starting point of the Boltzmann formalism is the
non-equilibrium distribution function in real and reciprocal space $f_{\sigma
}\left(  \mathbf{k},\mathbf{r}\right)  $ that in equilibrium reduces to
$f_{BE}^{\left(  0\right)  }\left(  \varepsilon_{\mathbf{k}\sigma}+\hbar\gamma
H_{0}\right)  .$ In the steady state, constant relaxation time $\tau_{r},$ and
one spatial dimension $x,$ the distribution is modified by a field or
temperature gradient as
\begin{align}
\triangle f_{\sigma}\left(  k,x\right)   &  =\tau_{r}\frac{\partial
\varepsilon_{k\sigma}}{\hbar\partial k}\frac{\partial f_{\sigma}\left(
k,x\right)  }{\partial x}\nonumber\\
&  =-\tau_{r}\frac{\partial\varepsilon_{k\sigma}}{\hbar\partial k}\left(
\frac{\hbar\gamma\partial H(x)-\partial\mu(x)}{k_{B}T}-\frac{\varepsilon
_{k\sigma}+\hbar\gamma H_{0}}{\hbar k_{B}T^{2}}\partial T\right) \nonumber\\
&  \times\left(  f_{BE}^{\left(  0\right)  }\right)  ^{2}e^{\frac
{\varepsilon_{k\sigma}+\hbar\gamma H_{0}}{k_{B}T}},
\end{align}
where we used the expansion Eq. (\ref{fBEexp}). Here we assume that the
relaxation time $\tau_{r}$ that includes all scattering processes is shorter
than $\tau_{m}$ in the spin diffusion equation.\ 

The magnon spin current in the magnon-polaron system for constant gradients
then, for example, reads%
\begin{equation}
-j_{m}=\sum_{k\sigma}\triangle m_{k\sigma}\triangle f_{\sigma}\left(
k\right)  =\sigma_{m}\left(  \partial H-\frac{\partial\mu}{\hbar\gamma
}\right)  -\sigma_{m}S_{m}\partial T,
\end{equation}
which leads to the conductivity%
\begin{equation}
\sigma_{m}=\frac{\tau_{r}}{k_{B}T}\sum_{k\sigma}\triangle m_{k\sigma}\left(
\frac{\partial\varepsilon_{k\sigma}}{\hbar\partial k}f_{BE}^{\left(  0\right)
}\right)  ^{2}e^{\frac{\varepsilon_{k\sigma}+\hbar\gamma H_{0}}{k_{B}T}},
\end{equation}
where $\triangle m_{k\sigma}$ is the magnetic component in the magnon-polaron
wave with index $k\sigma,$ which follows from diagonalizing the magnetoelastic
Hamiltonian \cite{Flebus}.

\textit{FEs - }The transport coefficients in ferroelectrics can be computed
under the ferron approximation analogously ($E_{0}=0$)
\begin{equation}
\sigma_{p}=\frac{\tau_{r}}{k_{B}T}\sum_{k\sigma}\triangle p_{k\sigma}\left(
\frac{\partial\varepsilon_{k\sigma}}{\hbar\partial k}f_{BE}^{\left(  0\right)
}\right)  ^{2}e^{\frac{\varepsilon_{k\sigma}}{k_{B}T}}. \label{sigmap}%
\end{equation}
where $\triangle p_{k\sigma}$ is the polarization (change) of a phonon in
state $k\sigma$ introduced in Eqs. (\ref{DelP},\ref{DelF}). The transport
coefficients in the 1D model with polarization parallel to the gradients can
be obtained analytically in the high and low temperature limits
\cite{Bauer2021}.

\subsubsection{Ballistic transport}

\textit{Magnets - }The Landauer-B\"{u}ttiker scattering theory of transport is
designed to treat transport that is limited by geometry, for example, by point
contacts that are connected adiabatically to large thermodynamic reservoirs.
Eq. (\ref{Onsagerp}) then becomes
\begin{equation}
\left(
\begin{array}
[c]{c}%
-J_{m}\\
J_{q}%
\end{array}
\right)  =G_{m}\left(
\begin{array}
[c]{cc}%
1 & S_{m}\\
\Pi_{m} & K/G_{m}%
\end{array}
\right)  \left(
\begin{array}
[c]{c}%
\triangle H\\
-\triangle T
\end{array}
\right)  , \label{LB}%
\end{equation}
where the driving forces are now the field and temperature differences, $K$
and $G_{m}$ are heat and polarization conductances, respectively, and on the
l.h.s. we have currents rather than current densities. The transport
coefficients are governed by the matrix of transmission probabilities
$\mathbf{T}$ of magnons that propagate from the left to the right, which in
the absence of scattering becomes a unit matrix in the space of propagating
states. The magnon conductance of the spin chain in the low temperature limit
\cite{Meier}
\begin{equation}
G_{m}=\hbar\gamma^{2}f_{BE}^{\left(  0\right)  }\left(  \varepsilon
_{m}\right)  .
\end{equation}
$H_{\mathrm{eff}}$ in $\varepsilon_{m}=\hbar\gamma H_{\mathrm{eff}}$
represents applied or magnetic anisotropy fields that fix the direction of the
magnetic order. $G_{m}\sim e^{-\varepsilon_{m}/\left(  k_{B}T\right)  }$
vanishes exponentially when the temperature falls below the magnon gap.

\textit{FEs - }In ferroelectrics \textit{ }%
\begin{equation}
\left(
\begin{array}
[c]{c}%
-J_{p}\\
J_{q}%
\end{array}
\right)  =G_{p}\left(
\begin{array}
[c]{cc}%
1 & S_{p}\\
\Pi_{p} & K/G_{p}%
\end{array}
\right)  \left(
\begin{array}
[c]{c}%
\triangle E\\
-\triangle T
\end{array}
\right)  ,
\end{equation}
where, e.g., the polarization conductance for the dipolar chain
\cite{Tang2021}
\begin{equation}
G_{p}=\frac{1}{k_{B}T}\sum_{\sigma}\int\left(  \xi_{\mathbf{p}}f_{BE}^{\left(
0\right)  }\right)  ^{2}e^{\varepsilon/(k_{B}T)}\frac{{d\varepsilon}}{{h}}.
\end{equation}
$G_{p}$ may be compared with the conductivity $\sigma_{p}$ of a diffuse wire
(\ref{sigmap}). The latter differs by being proportional to the relaxation
time and the increased importance of the group velocities in the integral.

For perpendicular FE order only the high-frequency optical branch contributes
with%
\begin{equation}
\xi_{\mathbf{p}}=-\frac{\left(  \hbar Q\right)  ^{2}}{MP_{0}\varepsilon}\text{
for }\mathbf{p}\perp\mathbf{\hat{x},}%
\end{equation}
while for parallel polarization with $\varepsilon\geq0$
\begin{equation}
\xi_{\mathbf{p}}=-\frac{\varepsilon P_{0}}{4C\delta^{2}}\text{ for }%
\mathbf{p}\Vert\mathbf{\hat{x}.}%
\end{equation}
We see that the ferron conductance dramatically differs from the magnon
conductance by its strong dependence on the direction of the dipolar order.
When $\mathbf{p}\perp\mathbf{\hat{x}}$, $G_{p}$ vanishes exponentially with
temperature because only the high-frequency optical phonon branch contributes.
For $\mathbf{p}\Vert\mathbf{\hat{x},}$ however, polarization transport is
gapless and the ferron conductance scales linearly with low temperatures
$G_{p}\sim T\ $\cite{Tang2021}.

\section{Devices}

The theory can be tested by experiments with concrete devices fabricated from
different materials and flexible configurations. Magnonics has a great
advantage: magnon currents can be measured when injected into heavy metal
contacts by means of a transverse electromotive force induced by the inverse
spin Hall effect \cite{Azevedo,Saitoh,Uchida}.

We are not aware of a \textquotedblleft polarization Hall
effect\textquotedblright\ that could detect a ferron current. Nevertheless,
ferrons do cause observable effects. Here we introduce two simple devices that
are tailored to find evidence for a polarization current in the diffuse and
ballistic regimes.

\subsection{Planar capacitors}

A capacitor is a slab of a dielectric insulator between metal contacts. Taking
advantage of the spin Hall effects, films with one or two Pt contacts on YIG
have been investigated extensively. Electric charges and capacitances appear
to play only a minor role in magnonics, however.

Planar FE capacitors are common devices because the high dielectric constant
of FE's ensures high capacitances and the switchability of the FE order allows
their use in FE random access memory (FeRAM) cells. Metallic contacts screen
the FE surface charges in equilibrium and residual electric fields are small
\cite{Mehta}. A global temperature change modulates the equilibrium
polarization and generates thermovoltages, while applied voltage steps heat or
cool the FE spacer, governed by susceptibilities $\chi_{p}$ and $\chi_{T}$.

A temperature \textit{gradient }generates a directional heat and associated
polarization current, but because $\partial E=0$ in these devices, an applied
voltage difference does not. We expect that a polarization current injected
into a metal contact decays on a very short length scale without obvious
observable signatures. However, opaque contacts accumulate a polarization on
the length scale of the diffusion length $\ell_{p}$. In an open-circuit
configuration, a thermally excited polarization $p\left(  x\right)  $
generates a voltage over the contacts
\begin{equation}
\triangle V=-\int_{0}^{L}\frac{\triangle p\left(  x\right)  }{\epsilon
}dx=-\frac{\chi_{E}}{\epsilon P}\int_{0}^{L}\mu\left(  x\right)  dx,
\end{equation}
where $L$ is the thickness of the FE barrier. In a capacitor with one contact
interface opaque and another one transparent and in the limit $\ell_{p}\gg L$
becomes $\triangle V_{\mathrm{\max}}=-\chi_{E}LS\triangle T_{\mathrm{ext}%
}/\left(  2\epsilon_{0}\epsilon_{r}\right)  $. $\triangle V$ can be used by
inserting the capacitor into an electric circuit, analogous to pyroelectric
devices, but operating on temperature differences rather than global
temperature changes. The effects can be switched off at temperatures below the
optical phonon band by rotating the polarization, as discussed above.

Returning to magnonics: a temperature gradient over a capacitor with a
electrically insulating magnetic\ spacer generates a magnon accumulation and
stray magnetic fields, which to the best of our knowledge have not yet been
reported, however.

\subsection{Lateral structures}

More flexible than the planar capacitors are lateral structures on and of thin
films. Propagating magnons can be injected and detected by narrow microwave
striplines \cite{Wang}, while Pt contacts on YIG\ films allow the study of
diffuse magnon transport by injecting them by the spin Hall effect or Ohmic
heating and detection in another contact by the emf generated by the inverse
spin Hall effect from spin pumping \cite{Cornelissenexp}.

We envisage analogous experiments in gated thin FE films. Their polarization
can be perpendicular to the plane, either spontaneously or induced by gates.
The ferroelectricity in few-monolayers of van der Waals materials such as
bilayer boron nitride in an especially interesting phenomenon \cite{Zheng}. In
contrast to the planar capacitor, it should be easy to generate electric field
gradients and electrically controlled polarization currents. Thin films may be
nanostructured such that large islands are connected by point contacts or
strips over which the electric field and temperature drop selectively, see
Figure 1. The scattering theory of transport is applicable and may be
approximated by the single wire model introduced above. Additional local
electrostatic gates should allow rotation of the FE order in the constriction.
These structures are open to local probes that can measure, for instance, the
significant magnetic stray fields predicted for the electric dipolar current
through the constriction \cite{Tang2021}.

\begin{figure}[pth]
\begin{center}
\includegraphics[width=10cm]{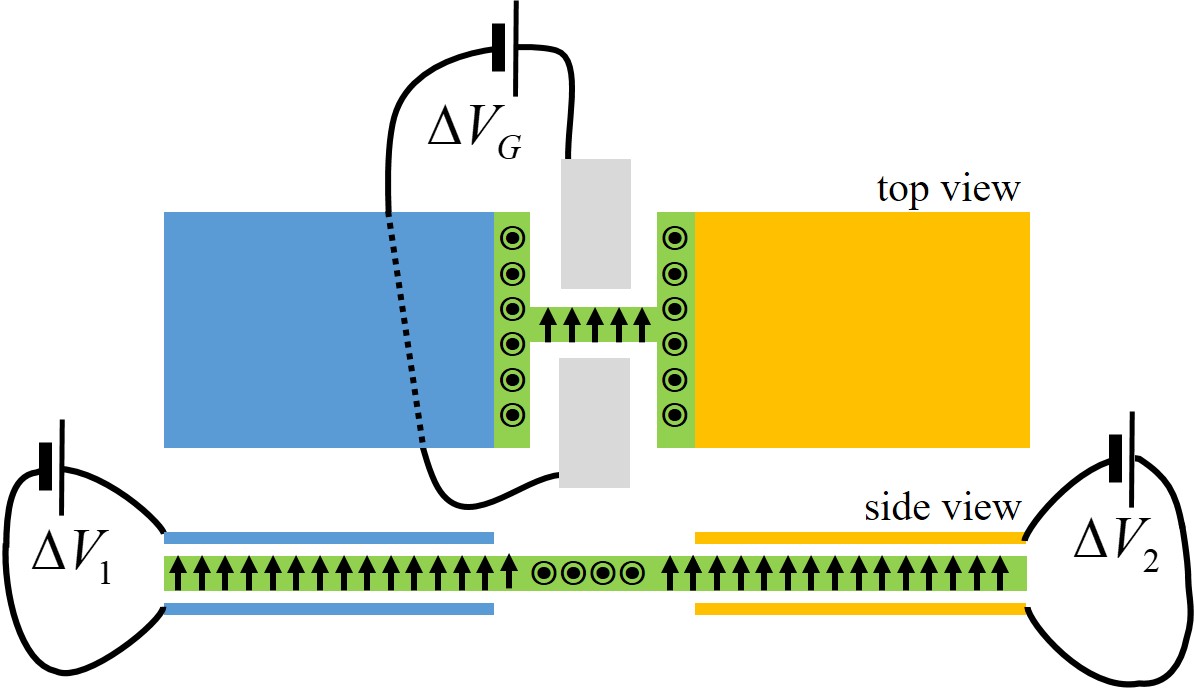}
\end{center}
\caption{Lateral polarization valve on a thin film FE. The two large FE pads
are thermodynamic reservoirs assumed to be at local thermal equilibrium, but
at variable electric fields and temperatures, controlled by metallic gates
with voltages $\Delta V_{1}$ and $\Delta V_{2}$ as well as temperatures
(indicated by the color). Here a voltage $\Delta V_{G}$ over side gates
controls the FE polarization direction angle with the sample plane.}%
\label{configuration}%
\end{figure}

\section{Summary and outlook}

Magnonics has developed into one of the most active fields in magnetism and
spintronics, while the field of \textquotedblleft ferronics\textquotedblright%
\ does not yet exist. Our models are primitive, some approximations are
untested, and essential parameters are not known. Experimental results are
necessary to guide better theoretical models and understanding. We hope that
our initial theoretical steps will motivate experimentalists with frontier
spirit to test our predictions and on the way discover completely new physics
and applicable new functionalities. Realistic lattice dynamic calculations of
the polarization amplitudes in ferroelectric materials can assess the accuracy
of the ferron approximation. The polarization relaxation time and diffusion
lengths are key parameters that have to be established with better accuracy.
We have only scratched the surface of the field, not addressing fascinating
extensions of ferronics that include antiferroelectrics, ferroelectric
textures, polar metals, and multiferroics.

\section*{Acknowledgements}

We thank K. Yamamoto, J. Heremans, and B. van Wees for helpful discussions.
G.B. acknowledges support by JSPS KAKENHI Grant No. 19H00645. R.\thinspace I.
and K.\thinspace U. are supported by JSPS KAKENHI Grant No. 20H02609, JST
CREST \textquotedblleft Creation of Innovative Core Technologies for
Nano-Enabled Thermal Management\textquotedblright\ Grant No. JPMJCR17I1, and
the Canon Foundation.


\begin{thebibliography}{99}                                                                                               %


\bibitem {Rezende20}S.M. Rezende, \textit{Fundamentals of Magnonics} (Springer
Nature, Heidelberg, 2020) ISBN 978-3-030-41317-0 .

\bibitem {Rezende67}S.M. Rezende and F. R. Morgenthaler, Magnetoelastic Waves
in Pulsed Magnetic Fields, Appl. Phys. Lett. \textbf{11}, 24 (1967), DOI:
10.1063/1.1754945 .

\bibitem {Rezende21}S. M. Rezende, D. S. Maior, O. Alves Santos, and J.
Holanda, Theory for phonon pumping by magnonic spin currents, Phys. Rev. B
\textbf{103}, 144430 (2021), DOI: 10.1103/PhysRevB.103.144430 .

\bibitem {Rezende14}S. M. Rezende, R. L. Rodr\'{\i}guez-Su\'{a}rez, R. O.
Cunha, A. R. Rodrigues, F. L. A. Machado, G. A. Fonseca Guerra, J. C. Lopez
Ortiz, and A. Azevedo, Magnon spin-current theory for the longitudinal
spin-Seebeck effect, Phys. Rev. B \textbf{89}, 014416 (2014), DOI:
10.1103/PhysRevB.89.014416 .

\bibitem {Bauer2021}G.E.W. Bauer, R. Iguchi, and K. Uchida, Theory of
Transport in Ferroelectric Capacitors, Phys. Rev. Lett. \textbf{126}, 187603
(1-4) (2021), DOI: 10.1103/PhysRevLett.126.187603

\bibitem {Tang2021}P. Tang, R. Iguchi, K. Uchida and G.E.W. Bauer, The
Ferroelectric Point Contact, arXiv:2105.14791 .

\bibitem {Ampere}A.-M. Amp\`{e}re, La d\'{e}termination de la formule qui
repres\'{e}nte l'action mutuelle de deux portions infiniment petites de
conducteur Volta\"{\i}ques, Ann. Chem. Phys. \textbf{20}, 398 (1922).

\bibitem {Gilbert}W. Gilbert, De Magnete (1700),

\bibitem {Spaldin}N.A. Spaldin, Topics Appl. Phys. \textbf{105}, 175 (2007).

\bibitem {Chandra07}P. Chandra and P. B. Littlewood, Topics Appl. Physics
\textbf{105}, 69 (2007).

\bibitem {Cornelissen}L.J. Cornelissen, K.J.H. Peters, G.E.W. Bauer, R.A.
Duine, and B.J. van Wees, Phys. Rev. B \textbf{94}. 014412 (2016).

\bibitem {Barker}J. Barker and G.E.W. Bauer, Semiquantum thermodynamics of
complex ferrimagnets, Phys. Rev. B \textbf{100}, 140401(R) (2019), DOI: 10.1103/PhysRevB.100.140401.

\bibitem {Berry}N.A. Spaldin, A beginner's guide to the modern theory of
polarization, J. Sol. St. Chem. \textbf{195, }2 (2012), DOI:
10.1016/j.jssc.2012.05.010 .

\bibitem {Flebus}B. Flebus, K. Shen, T. Kikkawa, K. Uchida, Z. Qiu, E. Saitoh,
R.A. Duine, and G. E. W. Bauer, Phys. Rev. B \textbf{95}, 144420 (2017).

\bibitem {Bauer}G.E.W. Bauer, E. Saitoh, and B.J. van Wees, Spin
Caloritronics, Nat. Mat. \textbf{11}, 391 (2011).

\bibitem {Meier}F. Meier and D. Loss, Magnetization transport and quantized
spin conductance, Phys. Rev. Lett. \textbf{90}, 167204 (2003).

\bibitem {Elyasi}M. Elyasi and G. E. W. Bauer, Cryogenic spin Seebeck effect,
Phys. Rev. B \textbf{103}, 054436 (2021), DOI: 10.1103/PhysRevB.101.054402.

\bibitem {Duine}R. A. Duine, A. Brataas, S. A. Bender, and Y. Tserkovnyak, In:
\textit{Universal Themes of Bose-Einstein Condensation}, edited by N.
Proukakis, D, Snoke, and P. Littlewood (Cambridge University Press, 2017), arXiv:1505.01329v1.

\bibitem {Azevedo}A. Azevedo, L.H. Vilela Le\~{a}o, R. L. Rodriguez-Suarez, A.
B. Oliveira, and S. M. Rezende, dc effect in ferromagnetic resonance: Evidence
of the spin-pumping effect?, J. Appl. Phys. \textbf{97}, 10C715 (2005), DOI: 10.1063/1.1855251.

\bibitem {Saitoh}E. Saitoh, M. Ueda, and H. Miyajima, Conversion of spin
current into charge current at room temperature: Inverse spin-Hall effect,
Appl. Phys. Lett. \textbf{88}, 182509 (2006), DOI: 10.1063/1.2199473 .

\bibitem {Uchida}K. Uchida, J. Xiao, H. Adachi, J. Ohe, S. Takahashi, J. Ieda,
T. Ota, Y. Kajiwara, H. Umezawa, H. Kawai, G. E. W. Bauer, S. Maekawa, and E.
Saitoh, Spin Seebeck insulator, Nat. Mat. \textbf{9}, 894 (2010), DOI: 10.1038/nmat2856.

\bibitem {Mehta}R. R. Mehta, B. D. Silverman, and J. T. Jacobs, J. Appl. Phys.
\textbf{44}, 3379 (1973).

\bibitem {Wang}H. Wang, J. Chen, T. Yu, C. Liu, C. Guo, H. Jia, S. Liu, K.
Shen, T. Liu, J. Zhang, M. A. Cabero Z, Q. Song, S. Tu, M. Wu, X. Han, K. Xia,
D. Yu, G. E. W. Bauer and H. Yu, Nonreciprocal coherent coupling of
nanomagnets by exchange spin waves, Nano Research, s12274-020-3251-5 (2020),
DOI: 10.1007/s12274-020-3251-5.

\bibitem {Cornelissenexp}L.J. Cornelissen, J. Liu, R.A. Duine, J. Ben Youssef,
B.J. van Wees, Long-distance transport of magnon spin information in a
magnetic insulator at room temperature, Nat. Phys. \textbf{11}, 1022 (2015),
DOI: 10.1038/nphys3465 .

\bibitem {Zheng}Z. Zheng, Q. Ma, Z. Bi, S. de la Barrera, M.-H. Liu, N. Mao,
Y. Zhang, N. Kiper, K. Watanabe, T. Taniguchi, J. Kong, W. A. Tisdale, R.
Ashoori, N. Gedik, L. Fu, S.-Y. Xu \& P. Jarillo-Herrero, Unconventional
ferroelectricity in moir\'{e} heterostructures, Nature \textbf{588}, 71
(2020), DOI: 10.1038/s41586-020-2970-9.
\end{thebibliography}
\end{document}